\documentclass[sigconf,nonacm]{aamas} 

\usepackage{balance} 
\usepackage{comment}

\setcopyright{ifaamas}
\acmConference[AAMAS '25]{Proc.\@ of the 24th International Conference
on Autonomous Agents and Multiagent Systems (AAMAS 2025)}{May 19 -- 23, 2025}
{Detroit, Michigan, USA}{A.~El~Fallah~Seghrouchni, Y.~Vorobeychik, S.~Das, A.~Nowe (eds.)}
\copyrightyear{2025}
\acmYear{2025}
\acmDOI{}
\acmPrice{}
\acmISBN{}

\acmSubmissionID{582}

\title[Detecting Strategic Behavior]{A Machine Learning Approach to Detect Strategic Behavior from Large-Population Observational Data Applied to Game Mode Prediction on a Team-Based Video Game}

\author{Boshen Wang}
\affiliation{
  \institution{University of Michigan-Dearborn}
  \city{Dearborn, Michigan}
  \country{United States of America}}
\email{boshenw@umich.edu}

\author{Luis E. Ortiz}
\affiliation{
  \institution{University of Michigan-Dearborn}
  \city{Dearborn, Michigan}
  \country{United States of America}}
\email{leortiz@umich.edu}

\begin{abstract}
Modeling the strategic behavior of agents in a real-world multi-agent system using existing state-of-the-art computational game-theoretic tools can be a daunting task, especially when only the actions taken by the agents can be observed. 
Before attempting such a task, it would be useful to gain insight into whether or not agents are in fact acting strategically at all, from a game-theoretic perspective.
In this paper, we present an initial step toward addressing this problem by proposing a general approach based on machine learning fundamentals for detecting potentially strategic behavior. 
We instantiate the approach by applying state-of-the-art machine learning tools for model selection and performance evaluation of prediction models in the context of detecting the strategic behavior of players for game mode selection in the multiplayer online video game \textit{Heroes of the Storm}.
Specifically, as a baseline, we first train neural networks to predict players' game mode selections using only information about the state of the player themselves.
Then, we train a new set of neural networks using the same architectures, this time incorporating ``historical co-play'' features that encode players' past interactions with other players.
We find that including these new features led to statistically significant improvements in game mode prediction accuracy, providing a sufficiently strong signal that players indeed make decisions strategically, which justifies the development of more complex computational game-theoretic tools in the hope of improving modeling and predictive power. 
We discuss remaining research work about potential approaches to validate the effectiveness of this initial step to detect strategic behavior.

\end{abstract}


\newcommand{\BibTeX}{\rm B\kern-.05em{\sc i\kern-.025em b}\kern-.08em\TeX}

\begin{document}


\pagestyle{fancy}
\fancyhead{}


\maketitle 

\section{Introduction}



%

Game theory provides a natural framework for analyzing multi-agent systems \cite{multi_agent_systems_book}, by offering concepts and tools to understands how rational agents make decisions, both as individuals and as a collective.
One defining feature of game theory is the idea that agents act \textbf{strategically}, meaning their preferences for how to act depend on the potential actions of other agents.
However, traditional game theory makes several assumptions about the knowledge each agent possesses, which might not reflect the the particular reality of the specific real-world multi-agent system being modeled or studied.
For example, in game theory it is assumed that each agent has preferences attached to every possible combination of actions they and other agents take.
In a real-world system with potentially numerous agents and a wide range of actions for those agents to choose from, it is likely that agents having ``perfect information'' about the system and its participants may be an unreasonable assumption.

Agents are not the only ones who have imperfect information about the system they are in.
Observers of real-world multi-agent systems may only have access to the decisions the agents made, without insight into their preferences or reasoning.
More sophisticated game-theoretic models, such as Bayesian games \cite{game_theory_fudenberg}, address some of these issues by relaxing some of the assumptions related to information availability. 
However, without knowing for certain the manner in which agents make decisions, how can we be sure that they are indeed acting strategically?
If, in truth, agents act independent of each other's decisions, then modeling the system using something like a Bayesian game, which carries its own technical and computational limitations, might be unnecessary.

As a first step to model real-world multi-agent systems in a game-theoretic way, it may be useful to first establish that agents are indeed acting strategically, even when only their actions are observable.
The use of sophisticated game-theoretic models can be justified only if evidence of strategic behavior is found.
In this paper, we consider a real-world multi-agent system consisting of players of the multiplayer online battle arena (MOBA) game \textit{Heroes of the Storm} (HotS), and propose a machine learning approach \cite{aima} to detect evidence of strategic decision making based on players' observed behavior in selecting game modes to play.


In Section \ref{sec:related}, we discuss related work on detecting strategic behavior of agents, and work related to predicting behaviors in video game settings via machine learning.
Section \ref{sec:data} provides context on the HotS game setting, and details the data used in our machine learning experiments.
We explain our experimental design choices and overall approach in Section \ref{sec:approach}.
We then describe the experimental setup in Section \ref{sec:experiments}.
The results of the experiments are discussed in Section \ref{sec:experiment_results}.
Finally, we consider future work related to our findings in Section \ref{sec:future}, and recap our contributions in Section \ref{sec:contributions}.







\section{Related Work}
    \label{sec:related}
    We will first discuss related work on detecting or assessing strategic behavior in game-theoretic settings.
    \citet{detect_strategy_peer} designed a statistical test to detect strategic behavior in a certain type of game in which players were incentivized to rank the work of peers in a way that benefited their own ranking.
    Human participants in their controlled experiments were instructed to consider behaving strategically for their own gain.
    Of course, without explicit instruction, humans may not behave strategically, even in situations where doing so would be beneficial.
    Along those lines, \citet{strategic_quotient} developed a quantitative assessment of human strategic ability called the \textit{strategic quotient}, which attempts to measure the ability of a person to reason about the potential actions of others.
    Using machine learning techniques to predict strategic behavior in games with potentially irrational human players was explored by \citet{deep_learn_predict_nf}.
    Their goal was to use a deep neural network to learn predictive factors that lead to irrational behavior, without the use of expert knowledge contributions such as strategic quotient.

    A crucial difference between the above works and the work presented in this paper was the availability of payoff information to both the players and the observers.
    In the above works, observations from experiments with humans participating in strategic games were obtained and studied, but said human participants, along with the observers, were all aware of the payoffs for every player in every outcome.
    This was in contrast to our HotS observational data, where only the actions of the players could be observed.
    Without knowledge of the players' payoffs or preferences, we did not have direct evidence that any player's payoffs depended on the actions of others, meaning there was no guarantee that any strategic behavior occurred.

    
    Our experimental approach, which is detailed in Section \ref{sec:approach}, shares many similarities with a study by \citet{chic_or_social}, which also attempted to find signs of strategic behavior in a real-world setting with only observable actions.
    In the study, images were obtained from a social media community focused on sharing fashion.
    Using computer vision and machine learning tools, the popularity of the fashion images were predicted based on the content of the image and social information about the image's poster.
    By comparing prediction accuracy of models trained only on image content features, models trained only on social features, and models trained using both feature sets, it was determined that social features were the most important predictive factor for popularity.
    By establishing that predictive models of image popularity only need to consider social factors of the image to perform well, it could be argued that users also mostly look at social factors rather than image content.
    Users deciding to ``like'' certain images because of social factors can be thought of as a kind of strategic behavior; the user's own preferences about what kind of image content they like were influenced or overridden by what other users have decided to like.
    
    Machine learning has been extensively applied to video game settings to make predictions based on observed human behavior.
    Oftentimes the research goal was to predict actions that players take while playing the game.
    Some examples of this include \citet{predict_hit_miss_fps}, who predicted if a shot taken in a shooting game will hit or miss, and \citet{predict_build_sc}, who predicted the next ``build'' action a player will take in a strategy game.
    The latter study was especially relevant because one of the data inputs to the action prediction model, a neural network, was a partial view of what the player's opponent was doing, based on information collected during the match.
    If the neural network, trained on human behavior in prior matches, predicted different actions due to the partially observed actions of the opponent, then that could be evidence of strategic behavior exhibited by the observed human players.
    
    Using observed behaviors in MOBA games in particular to make predictions or classifications is also common.
    For instance, \citet{win_predict_dota} and \citet{win_predict_league_rnn} aimed to predict the winner of a MOBA match using information both collected at the beginning of the match and collected as the match was played out.
    \citet{roles_class_dota} classified the ``role'' that players take on in a MOBA game, based on their behavior during the course of a match.
    Finally, \cite{draft_predict_hots} also performed a study using HotS data in order to predict character choices that a team of players would make.

    Somewhat surprisingly, to our knowledge there have been no studies on predicting the game mode selection of video game players.
    This can likely be partially attributed to the fact that many video games do not feature multiple game modes to choose from.
    Still, our game mode prediction experiments can be considered novel contributions, even if they were only used as a means to an end.


\section{Domain and Data}
    \label{sec:data}
    HotS is a team-based MOBA game developed by Blizzard Entertainment \cite{hots_blizzard}. 
    In HotS, players play matches in which they control a character (a hero) of their choosing and work together with teammates to battle players on the opposing team.
    The team that manages to destroy the other team's main base first wins the match.
    
    The data used in this paper was sourced from HotsAPI.net \cite{hotsapi_patreon}, now defunct. 
    HotsAPI.net was a data collection website run by members of the HotS player community.
    The site offered a way for players to voluntarily upload HotS match replay data automatically to the cloud.
    These replay data, which recorded details about the actions that players took during a HotS match, was then aggregated by other websites in the HotS community for informational purposes.
    
    In addition to its repository of replay files, HotSAPI also offered a database of summary data based on the replays.
    The summary data included match details such as the players participating on each team, the game mode of the match, and the outcome of the match.
    The summary database contained 18.2 million records belonging to 5.2 million unique player accounts, spanning four years from 2015 to 2019, with each record corresponding to one HotS match.
    For our experiments, we used this summary data to create a re-organized table in which each record corresponded to a player ID instead.
    Each match had exactly ten player IDs associated with it, so the re-organized data contained 182 million records total.

    Although we would like to use as much data as possible, practical concerns forced us to use only a subset.
    Since our goal was to study potential strategic interactions between players occurring before each match, we believed that match records for players with relatively few matches played ($<100$) were safe to discard.
    Players with many matches under their belt would have the most opportunities to engage in strategic decision making with other players.
    Another consideration was game server region.
    A player in one region cannot participate in a match hosted in a different region's server, meaning strategic interactions across different regions was not possible.
    Therefore, to further pare down the data, only records for players in the North America region were kept.
    After these cuts, about 61 million records remained.

\section{Approach Motivation}
    \label{sec:approach}
    In this section we discuss some of the challenges that accompany strategic behavior modeling in multi-agent systems, why HotS players could be modeled as strategic agents of a multi-agent system, and how we planned to detect strategic behavior using observational data.

    First we need to define a few terms from \textbf{game theory} \cite{game_theory_fudenberg}. 
    In game theory, a \textbf{game} is a mathematical model of a system in which at least two \textbf{players} or agents choose between one of several \textbf{actions} or strategies.
    Those actions have consequences for all players, referred to as \textbf{payoffs} or rewards.
    The combination of every player's chosen action is a \textbf{joint action}.
    In this paper we only consider \textbf{simultaneous} games, in which all players decide on an action at the same time without knowing exactly what other players have decided.
    In a simultaneous game, the outcomes of the game can be represented by a \textbf{payoff matrix}, which lists the payoff of every possible combination of player and action, organized in tabular form.
    This game representation is also referred to as \textbf{normal form}.

    Although our long term goal was to study strategic behaviors exhibited by agents in the multi-agent system consisting of online HotS players, modeling those behaviors presented significant technical challenges.
    For one, the number of joint actions or outcomes in the system is exponential in the number of players.
    In principle, if the system has $n$ players, and each player has $m$ actions to choose from, then the number of possible outcomes is $m^n$.
    A model that considers every possible outcome would quickly suffer from combinatorial explosion, making the model impractical.
    Another challenge was that only the actions taken by the players, not their payoffs, could be observed from the available HotS data.
    If a player's action selection was strategic, then their payoff for selecting a certain action should have been different depending on what actions other players had already selected.
    On the other hand, if a player's action selection was done independently, then their payoff for selecting some action should have been the same no matter what other players choose.
    Since we had no direct knowledge of the payoff matrix for any of the observed HotS players, we could not say for sure how many players were behaving strategically, or to what extent their decision making was strategic.

    Uncertainty about the players' payoffs or preferences begged the question, "Were there actually any opportunities for players to behave strategically in the system of HotS players?"
    We believed that players did indeed have the opportunity to act strategically when selecting a \textbf{game mode} to play. 
    Typically, when a player wishes to play a HotS match, they select a game mode based on their own preferences, and the game system then automatically finds teammates and opponents for them to play with.
    As an option, players are given the opportunity to form a ``party'' with up to four other willing players of their choosing, before deciding on a game mode.
    Players in a party have the benefit of always being assigned as teammates of each other by the game system, so long as they remain in the party together.
    However, in order to actually start a match, all players in the party must agree to play the same game mode.
    This restriction is enforced by the game software.
    Even though each player may have their own preferred game mode, they may decide to simply follow the game mode selections of their fellow party members, in spite of their own preferences.
    
    The situation where party members must agree on a game mode in order to play a HotS match is an instance of a coordination game from game theory \cite{coordination_games}. 
    In a coordination game, rational players who may prefer taking different actions will nonetheless all select the same action, because the payoff for agreeing (even on a less-preferred action) is higher than disagreeing.
    An exception to this is when a player in a HotS party refuses to coordinate their game mode selection with other party members, because their payoff for agreeing with the majority's choice is actually lower than their preferred choice. 
    In that case, the dissenting player will simply choose to leave the party, thus allowing themselves to once again pick their own preferred game mode.
    The remaining players in the party can then re-engage in a coordination game and finally agree on a game mode.

    Unfortunately, the data did not indicate if players in a match were actually in a party.
    Given that lack of information, how could we be sure which players, if any, actually behaved according to the above description?
    Suppose that all players actually chose game modes independently, and suppose that we used some machine learning algorithm to predict how each player selects game modes.
    Since selections were assumed to be independent, the learning algorithm would only need to have access to each player's state at the time of their next game mode selection.
    Any information about how other players selected game modes would be irrelevant, and including that information should have no impact on prediction accuracy.
    Suppose next that some players actually did act strategically, and chose game modes based on what other players had selected, to some extent.
    If that was the case, then knowing what game mode other players chose should improve prediction accuracy even with the same learning algorithm and model, simply due to more relevant data features being available.
    Using this idea, we could find evidence of strategic behavior without knowing the details of how players made decisions or valued different outcomes.

    Our experimental approach to detecting strategic behavior in HotS was as follows.
    For each player in a certain subset of the player population, we trained ``baseline'' predictors---specifically feed-forward \emph{artificial-neural-network (ANN)} classifiers~\cite{deep_learning_book} in our experiments---to perform game mode prediction, using only player state features.
    Then we trained new (ANN) predictors for the same task on the same subset of players, this time incorporating ``historical co-play'' features that encode how often the player's choice of game mode aligned with other players' choices.
    We then assessed if adding those historical co-play features led to significant improvements in average game mode prediction quality (in our case, classification accuracy), by evaluating and comparing the performance of the new predictors with the combined features with that of the baseline.
    We also trained one more set of predictors (also ANNs), this time using only the historical co-play features and none of the player state features, and assessed if those predictors ever led to improvements over the baseline or ``combined'' predictors networks.

    It should be noted that our goal was not necessarily to learn how to make game mode predictions as accurately as possible.
    Instead, the goal was to see if significant relative improvements could be made to the prediction model by simply including additional data features encoding the behavior of other agents.
    We chose to use a feed-forward ANNs for a few reasons: 1) there are numerous off-the-shelf tools to implement and train them, 2) shallow network architectures require relatively low amounts of data to train, and 3) they offer good performance in a wide variety of tasks.
    Although a more interpretable model such as a decision tree could have been used, we felt that the general purpose power of ANNs could be relied on to detect any signals of strategic behavior present in the historical co-play features.
    
    We believe that the machine-learning approach we outlined can be generalized to other settings in which only the actions of agents in a multi-agent system can be observed.
    Comparing the power of \textit{any} predictive model trained only on agent state features with the same model trained on both agent state features and features encoding the behavior of others should indicate if strategic behavior was present in the observations.
    It should be noted that introducing additional data features is guaranteed to improve training accuracy in many machine learning algorithms, so it is important to base evaluations on a hold-out or test data set~\cite{ml_textbook_murphy,learning_from_data_book}.

\section{Experiments}
    \label{sec:experiments}
    This section explains the design and setup of our strategic behavior detection experiments.
    For the baseline experiments, we limited the scope to predicting the game mode of each match played by certain chosen players, using only features describing the state of the player.
    Each chosen player had a neural network trained to predict the game mode selected for each of that player's matches.
    The set of chosen players $S$ was formed by simply selecting the $|S| \in \{1000, 2000, 4000\}$ players who had participated in the most matches.
    This was done for practical reasons; it would be infeasible to train a model for every player in the entire population, and the players in $S$ would also have the most records to train with in the data.
    Each network used the same feed-forward architecture and had the same features in their training data.
    The game mode prediction (classification) accuracy of each trained network was averaged together to obtained a final average prediction accuracy.

    For the ``historical co-play only'' experiments, we again trained a neural network for each player in $S$ which predicted the selected game mode for matches played by that player.
    This time, we only used features that encode past interactions with other players in $S$.
    The network architecture for the networks remained unchanged from the baseline experiments.
    Afterwards, the game prediction accuracy was again averaged across each of the trained networks, then compared with the baseline average accuracy.
    Finally, we repeated the process again for the ``combined'' experiments, where we used both player state features and historical co-play features.
    The three sets of experiments were re-run for different values of $n$ and different network architectures.

    
    In Section \ref{sec:features}, we describe the data features used for training the neural network prediction models in both the baseline and strategic detection experiments.
    Then, we discuss the network architectures of the neural networks, along with training details, in Section \ref{sec:network_arch}.

    \subsection{Data Features}
        \label{sec:features}
        The data used as input into the neural networks was sourced from the HotS summary database mentioned in Section \ref{sec:data}.
        Each record represented one match played by the current player in $S$.
        The target variable was the player's selected game mode for the match, which could take one of six distinct labels.
        In the baseline experiments, only features related to the state of the current player at the beginning of their match were used for training.
        The categorical features were encoded using a one-hot encoding, and the numerical features were normalized using z-score normalization.
        These input features are listed below, along with their rationale for inclusion:
        \begin{itemize}
            \item \textit{Hero Choice (Categorical)}: Some heroes are more useful in certain game modes, so the player's hero choice may give an indication of their chosen mode.
            \item \textit{Hour of Day (Categorical)}: The player may prefer certain game modes depending on the time of day.
            \item \textit{Day of Week (Categorical)}: The player may play certain game modes more often during certain days, like weekends.
            \item \textit{Game Date (Numerical)}: Due to game development updates, some game modes were not available until later dates.
            \item \textit{Arrival Time (Numerical)}: If a player had not played a match in a long time, then they may opt to play a casual game mode to ``warm up,'' before moving on to a more competitive game mode.
        \end{itemize}


        To detect strategic decision making by players with regards to picking a game mode, it would be tempting to simply add co-play features that indicate the presence of other players as teammates in the match, perhaps using a multi-hot encoding scheme.
        Specifically, we could add a feature column for each player in the player population, and set its value to 1 if the player corresponding to that feature column was present in the match, and 0 otherwise.
        If a player's game mode selection was influenced by certain other players' preferences, then the neural networks should be able to use the added features, in conjunction with the features used in the baseline experiments, to make more accurate game mode predictions.
        
        Unfortunately, there were two major issues with this approach.
        First, hundreds of thousands of unique players were present in the data.
        Adding a co-play feature corresponding to each player would add a prohibitive amount of overhead to populate the features and to initialize the neural networks.
        This drawback was addressed by adding co-play feature columns corresponding only to the players in $S$, rather than all players.

        Second, in HotS, a player chooses a game mode first before being assigned teammates.
        This fact means that we could not use information about the presence of other players to predict the game mode, because that information came from after the match began.
        The exception to this is when a player chooses to form a party of teammates.
        In that scenario, other players are deliberately invited to the party first, then a game mode is chosen for all of them to play together.
        Because the data did not indicate if players were in a party or not, we were forced to assume that the identity of the other players in the match were not known at the time of game mode selection.

        As an alternative, we instead used the proportion of matches the other players had played with the current player \textit{in the past} as teammates, up to when the current match occurred.
        We refer to these as ``historical co-play'' features.
        The idea was that if a player started playing many matches with certain other players, it could indicate a change in game mode preference for the current player.
        These proportions were not directly recorded in the data, but were computed by looping through the current player's match records and tracking the number of times other players in $S$ appeared in prior matches.



    \subsection{Network Architecture and Training}
        \label{sec:network_arch}
        For our experiments, feed-forward neural networks with a single hidden layer were used.
        The input layer contained the encoded features described in Section \ref{sec:features}.
        It varied in size from 120 for the baseline experiments, up to 1,400 neurons in combined experiments, depending on the size of $S$.
        If an input was found to have the same value in every record, then it was removed from the network before training.
        Networks with different hidden layer sizes $m \in \{10, 40, 160, 640, 1280\}$ were evaluated.
        The hidden layer neurons used the \textit{ReLU} activation function. 
        We also evaluated a network architecture which did not have a hidden layer, i.e. a single-layer perceptron.
        The output layer used the \textit{softmax} activation function and consisted of six neurons corresponding to the six possible game modes players can pick.
        Each layer was fully connected with neighboring layers.
        The neural networks were implemented with the Tensorflow \cite{tensorflow} and Keras \cite{keras_github} Python libraries, using Keras's ``functional API.'' 

        All networks in all experiments used the same training hyperparameters.
        However, recall that each network was trained to predict game modes for one player only, so each network was initialized as a new instance of the currently used model.
        Neuron weights were initialized using Glorot initialization \cite{glorot}, and biases initialized with zeros.
        The Adam optimization algorithm was used with learning rate 0.001. 
        Since the networks were trained to perform multi-class classification, categorical cross-entropy was the loss function used during optimization.
        Categorical accuracy was used as the evaluation metric, by comparing the argmax of the output layer with the input's true class label.

        To begin the training process, all records in the data pertaining to matches played by the current player were copied into a new collection data structure.
        These records were then shuffled and split into training, validation, and test data sets, using a 90/10/10 split.
        Note that due to the shuffling prior to training, temporal dependencies on consecutive records (in ``real time'') were ignored.
        The predictions were done purely on a ``snapshot'' of the player's current state.

        Z-score normalization of numeric features was done by computing the mean and variance of the training data only.
        The validation data set was only used to evaluate prediction accuracy during the training process for informational purposes.
        Networks were trained for 5 epochs.
        The test data set was used to perform the final evaluation of the trained model, and was not accessed in any way during training.
        
        Due to the amount of time required to train a neural network for every player in $S$, only about 9\% of the input data records were loaded.
        As a consequence, the number of networks trained during each experiment was only approximately $|S|*.09$.
        This made the average accuracy obtained at the end of each experiment significantly weaker evidence of true performance.
        Note, however, that the historical co-play features could still be computed properly, since only the current player's match history data was required for them.
        Also note that during each experiment, the same 9\% subset of the data was loaded.
        As we will see in Section \ref{sec:experiment_results}, there were still enough data points to show statistically significant improvements in average game mode prediction accuracy when including historical co-play features.

\section{Results and Discussion}
    \label{sec:experiment_results}

    \begin{table*}[h!]
        \caption{Average accuracy of neural network game mode predictions}
        \begin{center}
            \begin{tabular}{ |c|c|c|c|c|c| } \toprule
                \hline
                $|S|$       & \# Models     & Hidden Layer  & Baseline          & Co-play Only              & Combined                          \\ 
                            & Trained       & Size          & Avg. Accuracy     & Avg. Accuracy             & Avg. Accuracy                     \\ \midrule
                \hline                                                                                                                 
                1000        & 89            & 0             & $.696 \pm .031$   & $.667 \pm .015$           & $.731 \pm .026$                   \\
                1000        & 89            & 10            & $.701 \pm .031$   & $.714 \pm .017$           & $.736 \pm .026$                   \\
                1000        & 89            & 40            & $.727 \pm .027$   & $.721 \pm .016$           & $\textit{\textbf{.764}} \pm .023$ \\
                1000        & 89            & 160           & $.742 \pm .026$   & $.717 \pm .018$           & $\textit{\textbf{.779}} \pm .022$ \\
                1000        & 89            & 640           & $.754 \pm .025$   & $.716 \pm .018$           & $.784 \pm .023$                   \\
                1000        & 89            & 1280          & $.751 \pm .026$   & $.720 \pm .017$           & $.778 \pm .023$                   \\
                \hline                             
                2000        & 182           & 0             & $.670 \pm .022$   & $.652 \pm .015$           & $\textit{\textbf{.719}} \pm .018$ \\
                2000        & 182           & 10            & $.683 \pm .022$   & $.705 \pm .019$           & $\textit{\textbf{.726}} \pm .018$ \\
                2000        & 182           & 40            & $.712 \pm .019$   & $.718 \pm .017$           & $\textit{\textbf{.756}} \pm .016$ \\
                2000        & 182           & 160           & $.728 \pm .018$   & $.720 \pm .016$           & $\textit{\textbf{.763}} \pm .016$ \\
                2000        & 182           & 640           & $.738 \pm .018$   & $.714 \pm .017$           & $\textit{\textbf{.772}} \pm .015$ \\
                2000        & 182           & 1280          & $.743 \pm .017$   & $.711 \pm .017$           & $\textit{\textbf{.772}} \pm .016$ \\
                \hline                             
                4000        & 344           & 0             & $.672 \pm .017$   & $.641 \pm .016$           & $\textit{\textbf{.714}} \pm .013$ \\
                4000        & 344           & 10            & $.683 \pm .016$   & $\textit{.717} \pm .018$  & $\textit{\textbf{.729}} \pm .014$ \\
                4000        & 344           & 40            & $.708 \pm .014$   & $.723 \pm .017$           & $\textit{\textbf{.746}} \pm .012$ \\
                4000        & 344           & 160           & $.727 \pm .013$   & $.725 \pm .017$           & $\textit{\textbf{.756}} \pm .012$ \\
                4000        & 344           & 640           & $.739 \pm .013$   & $.719 \pm .018$           & $\textit{\textbf{.761}} \pm .012$ \\
                4000        & 344           & 1280          & $.739 \pm .013$   & $.716 \pm .018$           & $\textit{\textbf{.761}} \pm .012$ \\
        
                \hline  
            \end{tabular}
        \end{center}
        \label{tab:accuracy_results}
    \end{table*}

    \begin{figure}
        \includegraphics[width=1.0\linewidth]{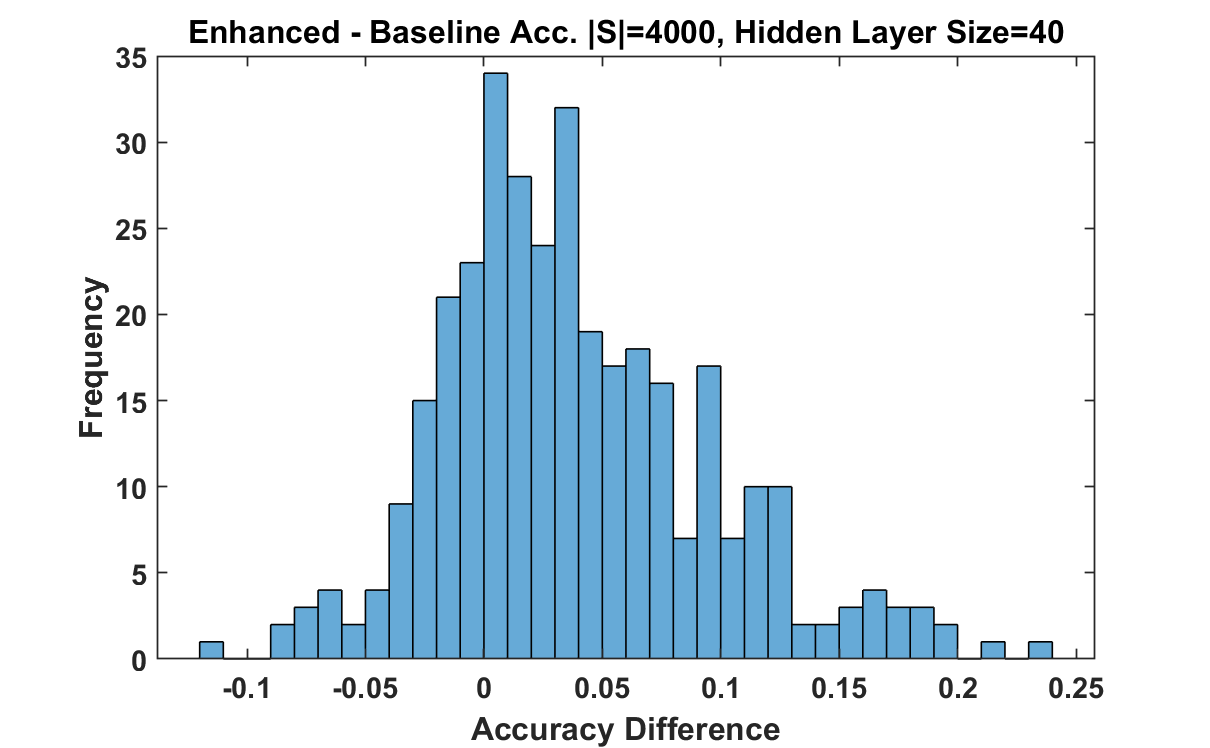}
        \caption{Histogram of accuracy difference between trained baseline and enhanced neural networks, for $|S|=4000$ and 40 hidden layer neurons.}
        \Description{Histogram of accuracy difference between trained baseline and enhanced neural networks, for $|S|=4000$ and 40 hidden layer neurons. The 0th, 25th, 50th, 75th, and 100th percentile value of the distribution are as follows: -0.11,0,0.03,0.07,0.23. }
        \label{fig:acc_diff}
    \end{figure}

    The resulting average game mode prediction accuracies and 95\% confidence intervals of the trained neural networks evaluated on a test data set for the baseline, historical co-play only, and combined experiments are presented in Table \ref{tab:accuracy_results}.

    Entries in which the hidden layer size is zero indicate that a single-layer perceptron was trained.
    In the ``Co-play Only'' column, an average accuracy value that in \textit{italics} indicates statistically significantly better accuracy than the corresponding baseline average accuracy.
    Likewise, in the ``Combined'' column, an average accuracy value in \textbf{\textit{bold and italics}} also indicates significantly better accuracy than the baseline.
    Statistical significance was established using a two-tailed z-test with $p = 0.05$.
    
    The main takeaway is that with all else being equal, combining historical co-play features with player state features led to better average prediction accuracy than the baseline in most situations, which was evidence that strategic behavior was taking place.
    When $|S| = 1000$, some improvement was still observed, but since the average accuracy was based on a relatively small sample size of 89 trained networks, the difference from the baseline result was not considered significant for every model.
    When $|S|$ was increased, more players in the loaded data files became members of $S$, leading to a larger sample size of trained network accuracies.
    With more samples, statistical significance was easier to establish.
    
    Increasing $|S|$ also led to more historical co-play features being used during training of the combined networks.
    However, it is not immediately clear if adding more historical co-play features actually improved average prediction accuracy in any way.
    Note that in the baseline experiments, as $|S|$ increases, the average prediction accuracy appears to decrease.
    This can be explained by considering that expanding $S$ does allow more data to be used, but that data is for players who have fewer matches recorded.
    Fewer recorded matches means less data to use for training, which would result in poorer prediction quality.
    We observe the same general trend of average accuracy decreasing as $|S|$ increases for the combined experiments.
    If adding additional historical co-play features actually improved game mode prediction accuracy, then its positive effect was small enough that it could not overcome the negative effect of training more networks for players who have less data.

    As expected, networks with fewer hidden layer neurons performed worse than networks with larger hidden layers in the three sets of experiments.
    The single-layer perceptron, despite being a relatively simple model, surprisingly did not perform much worse than a neural network with 10 hidden layer neurons in the baseline and combined experiments.
    The perceptron did under-perform in the ``co-play only'' experiments, however.
    Considering networks with a hidden layer, the largest boost in performance consistently came when increasing from 10 to 40 hidden layer neurons.
    Further increases to hidden layer size resulted in more modest accuracy improvements.
    An interesting note was that in the baseline experiments, the accuracy variance decreased when more hidden layer neurons were used, for the same $|S|$.
    Average test accuracy tended to stagnate or decrease once the hidden layer size reached 160 in the ``co-play only'' experiments, or 1280 in the baseline and combined experiments, signalling overfitting.

    Networks trained using only historical co-play features generally performed no better than the baseline of using only player state features.
    The one trial where ``co-play only'' significantly outperformed the baseline was when $|S|=4000$ and ten hidden layer neurons were used.
    Since the co-play features were generally useful when used in conjunction with player state features, but not on their own, we argue that although evidence of strategic behavior exists, not all players exhibited strategic behavior.
    If all players were acting strategically by coordinating their game mode selections with certain other players, then a network trained using only co-play features should have performed about as well as a network trained using both co-play and state features.
    
    Another observation of the ``co-play only'' experiments was that despite having up to $|S|$ features as input, a mere 10 to 40 hidden layer neurons were needed to make the best predictions for the given $|S|$.
    This possibly signals that only a relatively small number of the co-play features were necessary to make game mode predictions, meaning that if players did act strategically according to the behavior of others, only a small number of other players were actually considered.

    Although ``statistically significant improvement'' from the baseline is a promising quantitative result for the combined networks, a closer qualitative inspection of the individual experiment trials also tells a story.
    Recall that $S$ contained players with the most matches recorded in the data, so membership in the set did not change between experiments as long as $|S|$ remained constant.
    This meant that each player in $S$ who had a neural network trained to predict game modes during the baseline experiment also had a neural network trained for them during the combined experiments.
    Figure \ref{fig:acc_diff} plots the differences in individual categorical accuracy between the trained baseline neural networks and trained combined networks for a representative example, the case where $|S|=4000$ and 40 hidden layer neurons were used.
    Positive values mean that the accuracy of a trained neural network improved during the combined experiments, compared to its baseline network counterpart.
    
    We observe that most networks did not have much improved prediction accuracy, and some even had worse prediction accuracy.
    However, a subset of networks did see accuracy improvements of up to 23\%.
    For reference, the 25th, 50th, and 75th percentile accuracy difference values were 0.00, 0.03, and 0.07 respectively.
    The fact that not all networks saw an improvement when trained using historical co-play features combined with state features is more evidence that not all players in HotS behave strategically with regards to game mode selection.
    This makes sense intuitively, as not every player will engage in forming parties with other players and negotiate to coordinate game modes, as was described in Section \ref{sec:approach}.
    When not playing in a party, players have full freedom to choose game modes according to their own preferences, meaning no strategic interactions occur with any other players.
    However, including historical co-play features during training did make a positive impact for some players, which is evidence of strategic behavior occurring in that subset of players.
    


\section{Future Work}
    \label{sec:future}
    

    Although our proposed machine-learning approach was able to experimentally detect empirical evidence of strategic behavior in HotS players, we need additional experiments to better establish the generality and power of the proposed approach.
    For instance, the same experiments could be done on a synthetic data set, in which we finely control the strategic behavior (or lack thereof) of the observed agents.
    We could also try different predictive models to see if they improve detection power, and if certain models are better suited to detecting changes in behavior caused by strategic interaction. 
    There may also be alternative statistical or econometric tools to detect strategic behavior. For example, it may be possible to adapt tools to establish \emph{(Granger) causality}~\cite{granger_causality_review} on time series data to our context. It would also be interesting to compare such alternatives to the machine-learning approach proposed here.
    Of course, just as in the case of causality, theoretical limitations to detecting and differentiating truly game-theoretic strategic behavior from simply (probabilistic) correlated behavior are likely and establishing conditions under which such discrimination may be possible requires further research.

    Having now reasonable evidence of potential strategic behavior of players' game mode selection in the HotS dataset, we can address our initial goal of modeling the problem game theoretically. We could start by adapting and applying econometric approaches to infer large-scale Bayesian games from observational data to our setting~\cite{econometric_inf_Bayesian_games, linear_social_models}. Such econometric approaches typically come with strong theoretical identifiability guarantees that require equally strong assumptions on the underlying model and the generative process, and only hold \emph{asymptotically} (i.e., in the limit of infinite amount of data). Recently, alternative approaches have been proposed for learning games (e.g., network-structured compactly-representable games such as \emph{graphical games}~\cite{Kearns_2007}) from observational data that relax the strong modeling assumption and conditions on the generative process, and in some instances also provide finite sample complexity guarantees generally at the cost of lack of identifiability~\cite{local_agg_games,learning_gg_cond,learning_linear_inf_games,learning_sparse_gg,learning_tree_pot_games}. Applying those approaches to our setting will require significant research to extend them to handle Bayesian and stochastic games. One possibility is to formulate generative models of behavioral data from game theoretic models under the assumption that the time series of observed agent interactions is the results of the agents following some type of learning process of the type typically studied in the literature of \emph{learning in games}~\cite{fudenberg1998theory,Blum_Mansour_2007,NEURIPS2021_e85cc63b}, which is primarily interested in studying the behavior and properties by which players \emph{learn to play a game} (given known payoffs or payoff feedback), and design statistical machine-learning methods to infer the game parameters. A related potential avenue of research is to explore the application of algorithms for \emph{multi-agent inverse reinforcement learning}~\cite{yu2019multi,multi_agent_inverse_rl,10.1613/jair.1.12594,wu2023multiagent,goktas2024efficient} to our setting, which will require scaling those algorithms to effectively handle games with many players and noisy data in incomplete information, partially observable domains. Finally, synthetic experiments may help us study how the predictive power of detection using the proposed machine-learning approach correlates with the performance of the resulting game-theoretic models. For example, we could study whether, or to what degree, larger improvements in performance of the combined predictors relative to the baseline predictors translate to larger improvements on the quality of game-theoretic models over more standard probabilistic/stochastic models when learning from observed behavioral data, further motivating the use and applicability of game-theoretic models in real-world strategic domains.
    
\section{Contributions and Closing Remarks}
    \label{sec:contributions}
    Real-world multi-agent settings can present significant technical challenges for modeling the strategic behavior of agents in a game-theoretic framework. This is in large part because of the lack of knowledge about agents' payoffs.
    Uncertainty about the desire or ability for agents to act strategically necessitates a method to detect strategic behavior in the multi-agent system as a first step to determine whether considering game theoretic approach, which often requires additional modeling and computational complexity, would be appropriate.
    We proposed a general approach to detect strategic behavior in multi-agent systems in which only the actions of the agents are observable based on machine learning evaluation fundamentals.
    We specifically applied our approach using deep learning techniques to predict game mode selection behavior collected in a large and novel dataset of HotS match summary records, to illustrate the effectiveness of the approach in that particular domain.
    We believe that our results establishing with high statistically significant degree of confidence that performance of predictors that include other player historical choices over those that only account for individual player context/state information suggest that there is indeed strategic behavior in game mode selection.
    We also found that although strategic behavior played a role in the observed system, only a subset of players seem to actually engage strategically, while the rest seem to act independently.
    
\begin{acks}
This work was supported in part by NSF Award IIS-1907553.
\end{acks}
\bibliographystyle{ACM-Reference-Format} 
\bibliography{mode_prediction}


\end{document}